\documentclass[twocolumn,aps,prb]{revtex4-2}
\usepackage[pdftex]{graphicx}
\usepackage{dcolumn}
\usepackage{bm}
\usepackage{float}
\usepackage{hyperref}
\usepackage{epstopdf}
\usepackage{mathrsfs}
\usepackage{graphicx}
\usepackage{multirow}
\usepackage{amsmath}
\usepackage{amssymb}

\newcommand{\ket}[1]{\left| #1 \right>}
\newcommand{\bra}[1]{\left< #1 \right|}

\newcommand{\Fig}[1]{Fig.\,\ref{#1}}
\newcommand{\Eq}[1]{Eq.\,\eqref{#1}}

\hypersetup{
            colorlinks=true,
            linkcolor=blue,
            anchorcolor=blue,
            citecolor=blue}
\begin{document}

\title{Non-Markovian quantum Mpemba effect in strongly correlated quantum dots}
\author{YuanDong Wang$^{1}$}
\email{ydwang@cau.edu.cn}
\affiliation{
$^{1}$Department of Applied Physics, College of Science, China Agricultural University, Qinghua East Road, Beijing 100083, China.\\
 }

\begin{abstract}
Harnessing non-Markovian effects has emerged as a resource for quantum control, where a structured environment can act as a quantum memory. We investigate the quench dynamics from specific initial states to equilibrium steady states in strongly correlated quantum dot systems. The distance between quantum states is quantified using the Bures metric, which endows the space of reduced density matrices with a Riemannian geometric structure. Using the numerically exact hierarchical equations of motion (HEOM) method, we demonstrate a quantum Mpemba effect arising from non-Markovianity. This effect is characterized by a relaxation slowdown due to information backflow from the bath to the system, which induces a pronounced memory effect. We show that the emergence of the non-Markovian quantum Mpemba effect on the approach to a strongly correlated steady state is determined by the interplay between the initial-state-dependent non-Markovianity and the initial geometric distance between states. Our results underscore the critical role of memory effects in quantum quench dynamics and suggest new pathways for controlling anomalous relaxation in open quantum systems.
\end{abstract}

\maketitle

The Mpemba effect (ME)—a counterintuitive phenomenon where a system initially at a higher temperature can relax faster than one starting from a lower temperature—was first observed in water over half a century ago \cite{EBMpemba_1969}. While numerous mechanisms have been proposed to explain the ME \cite{VYNNYCKY20127297,VYNNYCKY2015243,MIRABEDIN2017219,10.1119/1.18059,C4CP03669G,https://doi.org/10.1002/crat.2170230702,PhysRevE.100.032103,10.1119/1.3490015,10.1119/1.2996187,burridge2016questioning}, a universal explanation remains elusive. The analogous quantum Mpemba effect (QME) occurs when a quantum system initially farther from equilibrium thermalizes more rapidly than one initially closer to it \cite{Ares2025,teza2025speedups}. This effect has been studied in both open quantum systems \cite{PhysRevLett.127.060401, PhysRevResearch.3.043108,e27060581} and isolated systems \cite{PhysRevB.100.125102,Ares2023,Murciano_2024,PhysRevB.110.085126}. For Markovian  open quantum systems, where thermal baths are assumed to be memoryless, the dynamics are governed by a Lindblad master equation. Within this framework, the QME can be characterized by the spectral properties of the Liouvillian: for specific initial conditions, the slowest decaying mode can be either eliminated (strong ME) \cite{PhysRevLett.127.060401} or suppressed (weak ME), leading to exponentially faster relaxation. This spectral origin is shared by classical Markovian systems described by a Fokker-Planck equation \cite{PhysRevX.9.021060,Kumar2020}. Recent experimental observations of the QME have been reported in several platforms \cite{PhysRevLett.133.010403,Zhang2025,PhysRevLett.133.010402,PhysRevLett.133.140405,liu2024quantum,turkeshi2024quantum}.

However, when environmental memory effects are significant, the dynamics of an open quantum system deviate from a dynamical semigroup, leading to the breakdown of the Markov approximation \cite{RevModPhys.88.021002}. This raises fundamental questions about whether non-Markovianity governs the quantum Mpemba effect (QME), particularly regarding how memory effects influence the distance between instantaneous states and the equilibrium state, and these questions remain unresolved \cite{PhysRevLett.134.220403}. Violations of the Markov approximation typically occur under strong system-environment coupling or low temperature conditions \cite{breuer2002theory}. In such regimes, the Lindblad master equation becomes invalid and spectral methods inapplicable, as the dynamics can no longer be described by simple exponential decay modes. In this Letter, we demonstrate that strong many-body correlations induce a distinct form of QME mediated by non-Markovianity, which transcends the conventional classification of strong and weak QME.

 A prominent example of strong correlation in open quantum systems is the Kondo effect \cite{10.1143/PTP.32.37,Michael_Pustilnik_2004,hewson1993kondo}, wherein a local magnetic moment from an isolated electron is screened by itinerant electrons via an antiferromagnetic interaction at low temperatures. The strong system-environment coupling inherent to this effect gives rise to non-Markovian signatures, such as the Kondo resonance \cite{PhysRevLett.111.086601, chan2024revealing,cao2024simulating}. Quantum dots (QDs) provide a versatile platform for engineering and probing the Kondo effect, which is theoretically captured by the Anderson impurity model (AIM) \cite{PhysRev.124.41}. The total Hamiltonian is given by  $H=H_{\text{dot}} +H_{\text{res}}+ H_{\text{res}}$.  The QD Hamiltonian is  $H_{\text{dot}}=\sum_{s=\uparrow,\downarrow} \epsilon_d d_{s}^\dagger d_s + U n_{\uparrow}n_{\downarrow}$,  with and $d_{s}^\dagger$ ($d_{s}$) creates (annihilates) a spin-$s$ electron on the dot level of energy $\varepsilon_d$.  And $U$ is the Coulomb energy whereas $n_{s}=d_{s}^\dagger d_{s}$. The electrodes are described by noninteracting electron reservoirs, $H_{\text{res}}= \sum_{\alpha ks}\epsilon_{\alpha ks}c_{\alpha ks}^\dagger c_{\alpha ks}$, where $\epsilon_{\alpha ks}$ is  the single particle energy level of the $k$ state with $s$  spin in the $\alpha$ lead, and $c_{\alpha ks}$  ($c_{\alpha ks}^\dagger$) is the corresponding annihilation (creation) operators, respectively. The dot-reservoir coupling is described by $H_{\text{coup}} = \sum_{\alpha k}t_{\alpha k}d_{s}^{\dagger}c_{\alpha ks}+\text{H.c.}$, with $t_{\alpha k}$ representing the tunneling amplitude  between the QD and the the reservoir. The hybridization function of the electrode assume  a Lorentzian form $J_{\alpha}(\omega) \equiv \pi \sum_{k}t_{\alpha k}t_{\alpha k}^{*}\delta(\omega - \epsilon_{\alpha k}) = \Gamma_{\alpha} W^2/[(\omega -\mu_\alpha)^2 + W^2]$, where $\Gamma_{\alpha}$ is the dot-reservoir coupling strength, $W$ is the bandwidth, and $\mu_\alpha$ is the chemical potential of the $\alpha$ reservoir.

However, the accurate and efficient characterization of many-body correlations and non-Markovian memory effects in open quantum systems remains a longstanding challenge. 
Several theoretical approaches grounded in non-perturbative quantum dissipation theories have been developed, including the stochastic equation of motion (SEOM)~\cite{10.1063/1.1647528,PhysRevLett.123.050601,PhysRevLett.88.170407,10.1063/1.4984260}, quantum state diffusion (QSD)~\cite{PhysRevA.58.1699,PhysRevLett.82.1801,PhysRevA.69.052115,PhysRevLett.105.240403,PhysRevLett.119.180401}, the hierarchy of stochastic pure states (HOPS)~\cite{PhysRevLett.113.150403}, and quantum Monte Carlo (QMC) methods~\cite{PhysRevLett.115.266802,PhysRevLett.130.186301}, among others. 
In this work, we employ the hierarchical equations of motion (HEOM) method~\cite{doi:10.1143/JPSJ.58.101,PhysRevA.41.6676,10.1063/5.0011599,10.1063/1.2938087}, which captures non-Markovian effects through memory kernels encoded in two-time correlation functions. 
The HEOM formalism describes the reduced dynamics of the system---here, impurity electrons or $f$ electrons---via the time evolution of the reduced density operator $\rho^{(0)}(t)$. 
The thermodynamic equilibrium of the $\alpha$-reservoir is characterized using the grand canonical ensemble. 
The time derivative of $\rho^{(0)}(t)$ couples to first-order auxiliary density operators (ADOs) $\rho^{(1)}(t)$, while higher-order ADOs $\rho^{(i)}(t)$ are hierarchically coupled to $\rho^{(i-1)}(t)$. 
This hierarchy can be compactly expressed as:
\begin{equation}
\begin{aligned}
\dot{\rho}^{(n)}_{j_1\cdots j_n}=&-\left( i\mathcal{L}_{\text{sys}} +\sum_r \gamma_{jr} \right)\rho^{(n)}_{j_1 \cdots j_n} -i \sum_j \mathcal{A}_{\bar{j}}\rho_{jj_1\cdots j_n}^{(n+1)}\\
&-i\sum_r (-1)^{n-r} \mathcal{C}_{j_r}\rho_{j_1\cdots j_{r-1}j_{r+1}\cdots j_n}^{(n-1)}.
\end{aligned}
\end{equation}
The system dynamics are governed by the Liouville superoperator $\mathcal{L}_{\text{sys}}[\cdot] = [H_{\text{sys}}, \cdot]$. 
The reduced density matrix (RDM) of the system, denoted $\rho^{(0)} \equiv \rho_S = \text{tr}_B[\rho_T]$, represents the system degrees of freedom, where $\rho_T$ is the total system-environment density matrix. 
The hierarchical structure comprises auxiliary density operators (ADOs) $\{\rho_{j_1\cdots j_n}^{(n)}; n=1,2,\ldots\}$ at the $n^{\text{th}}$ tier, each associated with a specific set of dissipation modes $\{j_1,\ldots,j_n\}$. 
A multicomponent index $j = \{\sigma, \alpha, \nu, p, s\}$ labels the principal dissipation modes. 
The superoperators $\mathcal{A}_{j}$ and $\mathcal{C}_{jr}$ are defined by their action on an arbitrary operator $O$ as $\mathcal{A}_j O = [d_{s}^{z}, O]$ and $\mathcal{C}_{j} O = \eta_j d_{s}^{z} O \pm \eta_{j}^{*} O d_{s}^{z}$, where $z = \pm$ and $\bar{z} = -z$. 
The parameters $\eta_{j}$ and $\gamma_{jr}$ originate from the parameterized hybridization function. 
Further details regarding the HEOM methodology can be found in Refs.~\cite{doi:10.1143/JPSJ.58.101,10.1063/1.2938087,PhysRevLett.109.266403,10.1063/1.4863379,https://doi.org/10.1002/wcms.1269}.

\begin{figure}[htbp]
\centering
\includegraphics [width=1 \columnwidth]{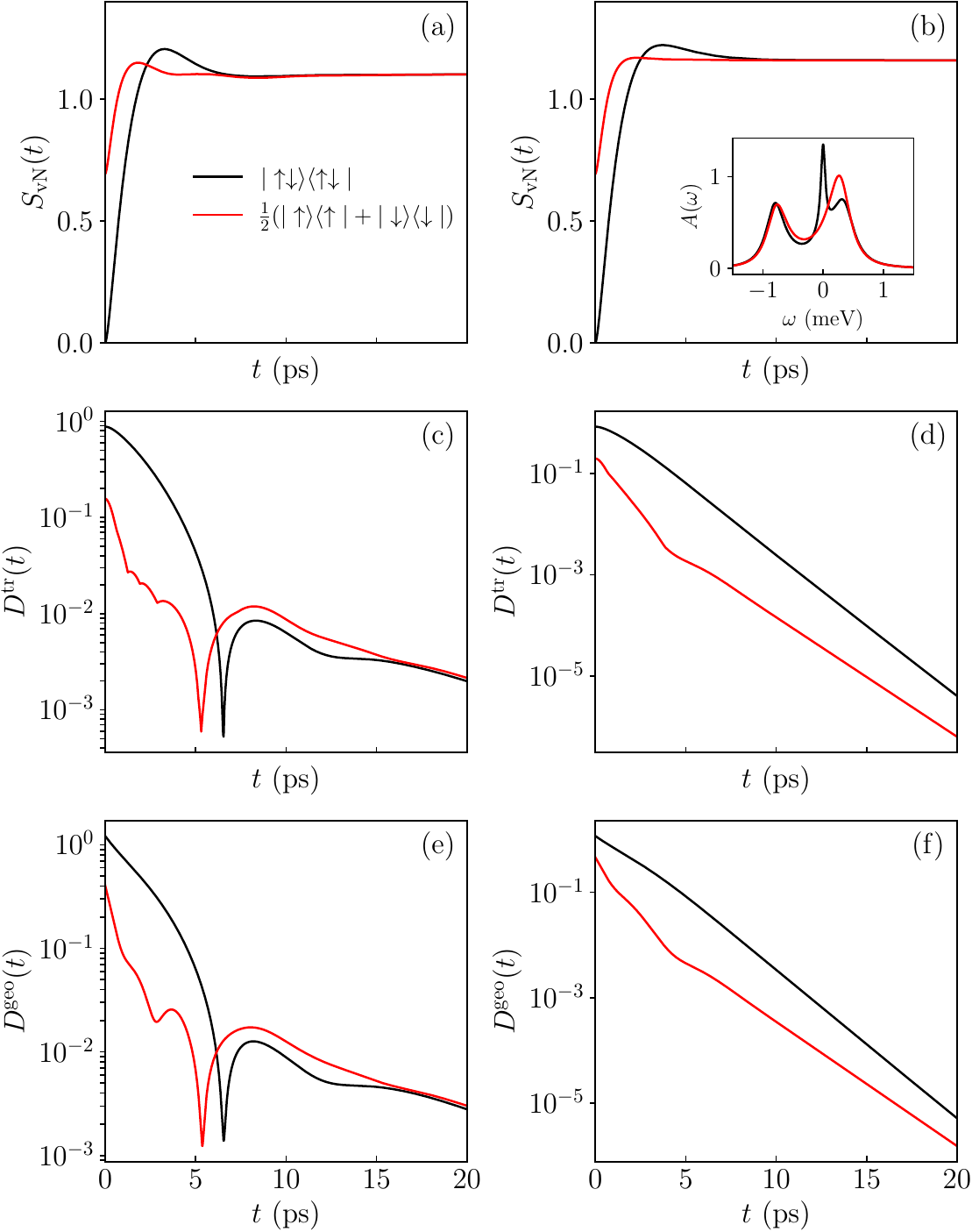}
\caption{Time-evolution of von Neumann entropy (a) and (b), trace distance (c) and (d), and geodesic distance (e) and (f) for initial states $\ket{\uparrow\downarrow}\bra{\downarrow\uparrow}$ and $\frac{1}{2}(\ket{\uparrow}\bra{\uparrow} + \ket{\downarrow}\bra{\downarrow})$. For (a), (c), (e) the temperature is $\tilde{T}=0.1$ while for (b), (d), (f) $\tilde{T}=1$. The inset in (e) is the spectral functions. Other parameters are set as $\epsilon_{d} = -0.7$, $U=1$,  $\Gamma=0.2$, $W=2$, in unit of meV.  }\label{fig1}
\end{figure}

Various distance measures quantify the deviation of a system from equilibrium. 
One such measure is the trace distance $D^{\text{tr}}(\rho,\sigma) = \frac{1}{2}\operatorname{Tr} \lvert \rho - \sigma \rvert$, which corresponds to the maximal probability of distinguishing $\rho$ and $\sigma$ in a single measurement~\cite{RevModPhys.88.021002}. 
A key property of the trace distance is its contractivity under completely positive trace-preserving (CPTP) maps: 
When the system and environment are initially uncorrelated ($\rho_{SB} = \rho_S \otimes \rho_B$), any quantum channel $\Phi$ satisfies $D^{\text{tr}}(\Phi \rho, \Phi \sigma) \leq D^{\text{tr}}(\rho, \sigma)$. 
Consequently, Markovian dynamics induce monotonic decrease of the trace distance. 
By contrast, non-Markovian dynamics permit temporary increases in $D^{\text{tr}}$ (for details see Appendix~\ref{appa}), signifying information backflow from the environment~\cite{PhysRevLett.103.210401,RevModPhys.88.021002}.

Although the trace distance serves as a hallmark of non-Markovianity, it lacks a direct geometric interpretation as a statistical distance. 
To address this limitation, one may introduce a Riemannian metric on the manifold of density matrices. 
The Uhlmann fidelity $\mathcal{F}(\rho,\sigma) = \operatorname{Tr} \sqrt{\rho^{1/2} \sigma \rho^{1/2}}$~\cite{uhlmann1976transition,nielsen2010quantum,petz2011introduction} quantifies the closeness of two density matrices and exhibits symmetry and unitary invariance. 
While fidelity itself is not a distance measure, it enables the definition of the Bures distance $D^{\text{B}}(\rho,\sigma) = \sqrt{2(1 - \mathcal{F}(\rho,\sigma))}$, which corresponds to the Euclidean distance between points on a unit sphere. 
The Bures distance between infinitesimally close density matrices yields the Bures metric~\cite{Hans-Jurgen-Sommers_2003}:
\begin{equation}\label{b-metric}
ds^{2} = [D^{\text{B}} (\rho,\rho+\delta\rho)]^2 = \frac{1}{2} \sum_{\nu\mu} \frac{\lvert \langle \nu | d\rho | \mu \rangle \rvert^2}{\rho_\nu + \rho_{\mu}},
\end{equation}
where $\{|\nu\rangle\}$ are the eigenstates of $\rho$ with eigenvalues $\rho_\nu$. 
The corresponding geodesic distance,
\begin{equation}
D^{\text{geo}}(\rho,\sigma) = 2 \arccos \mathcal{F}(\rho,\sigma),
\end{equation}
defines the Bures angle (or Bures length). For pure states, this expression reduces to the Fubini–Study distance.

For the trace distance, despite as a hallmark of non-Markovianity, it lacks a direct geometric meaning of statistical distance. On account of this, an alternative  option is to  settling a Riemannian metric on the manifold of density matrices.  The Uhlmann fidelity $\mathcal{F}(\rho,\sigma)=\text{Tr}\sqrt{\rho^{1/2}\sigma \rho^{1/2}}$ \citep{uhlmann1976transition, nielsen2010quantum, petz2011introduction}  measures how close two density matrices and is symmetric and invariant under unitary operations. The fidelity itself does not define a distance directly, but it allows us to define a proper metric, the so-called Bures distance $D^{\text{B}}(\rho,\sigma) =\sqrt{ 2(1-\mathcal{F}(\rho,\sigma))}$, which is the Euclidean distance between two points on a unit circle. The Bures distance between two infinitesimally close density gives the Bures metric \cite{Hans-Jurgen-Sommers_2003}: 
\begin{equation}\label{b-metric}
ds^{2} = [D^{\text{B}} (\rho,\rho+\delta\rho)]^2=\frac{1}{2}\sum_{\nu\mu}\frac{\lvert \bra{i} d \rho \ket{j}\rvert^2}{\rho_\nu + \rho_{\mu}}.
\end{equation} 
Via Bures metric, the geodesic distance is obtained as 
\begin{equation}
D^{\text{geo}}(\rho,\sigma)=2\arccos{\mathcal{F}(\rho,\sigma)},
\end{equation}
which is called the Bures angle or length, for pure states it reduces the Fubini–Study distance.

We now define the QME using the geodesic distance. 
Consider an initial state $\varrho \equiv \rho(t=0)$ that is suddenly coupled to the bath at $t=0$ (a quench). 
Let $\rho_{\text{ess}} \equiv \rho(t\rightarrow \infty)$ denote the equilibrium steady state. 
For two initial states $\varrho_{A}$ and $\varrho_{B}$, both uncorrelated with the bath, QME occurs when
\begin{equation}
\begin{aligned}
&D^{\text{geo}}(\varrho_{A},\rho_{\text{ess}}) > D^{\text{geo}}(\varrho_{B},\rho_{\text{ess}}), \\
\exists t_M > 0: \forall t > t_M: \quad &D^{\text{geo}}(\rho_{A}(t),\rho_{\text{ess}}) < D^{\text{geo}}(\rho_{B}(t),\rho_{\text{ess}}),
\end{aligned}
\end{equation}
where the subscript of $\rho(t)$ specifies the initial state. 
This condition implies that state $A$, initially farther from the ESS, reaches equilibrium faster than state $B$, which begins closer to the ESS.

 \begin{figure*}[htbp]
\centering
\includegraphics [width=2 \columnwidth]{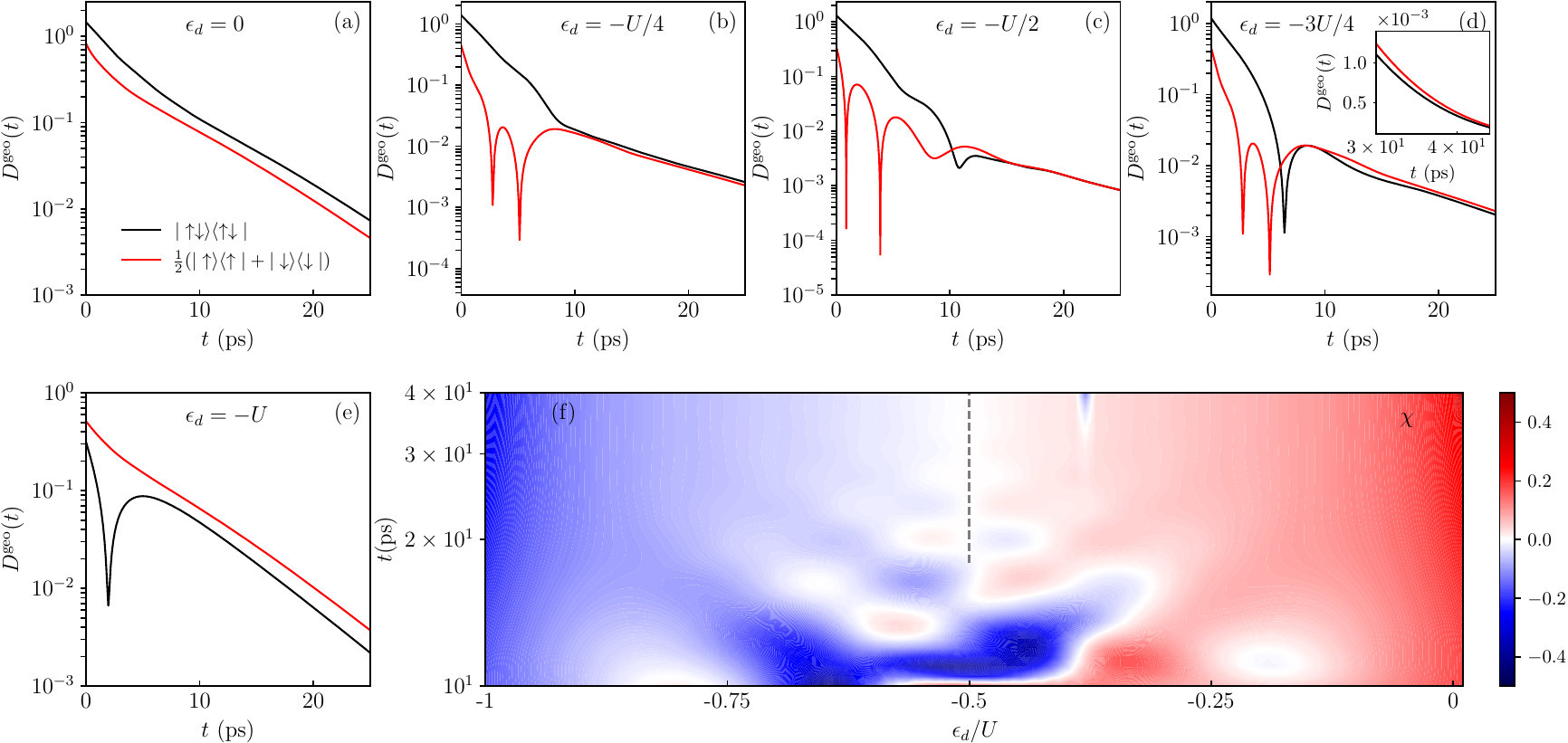}
\caption{(a)-(e) Geodesic distance evolution from initial states  $\ket{\uparrow\downarrow}\bra{\downarrow\uparrow}$ and $\frac{1}{2}(\ket{\uparrow}\bra{\uparrow} + \ket{\downarrow}\bra{\downarrow})$ for different single electron energy level $\epsilon_d$. The inset of (d) show the asymptotic behavior for long-time evolution. (f) The geodesic distance polarization $\chi$ versus time $t$ and $\epsilon_d$. For (a)-(f) temperature is set as $\tilde{T}=0.02$, other parameters are same as that in \Fig{fig1}.  }\label{fig2}
\end{figure*}

To isolate non-Markovian effects, the quantum dot (QD) and bath are prepared in an uncorrelated initial state (see Appendix~\ref{appa}). 
The QD Fock state basis consists of $\ket{0}$, $\ket{\uparrow}$, $\ket{\downarrow}$, and $\ket{\uparrow\downarrow}$. 
We consider two initial states: the doubly-occupied pure state $\varrho_{\text{db}} = \ket{\uparrow\downarrow}\bra{\uparrow\downarrow}$ and the singly-occupied mixed state $\varrho_{\text{sm}} = \frac{1}{2}(\ket{\uparrow}\bra{\uparrow} + \ket{\downarrow}\bra{\downarrow})$. 
System-environment entanglement is quantified by the von Neumann entropy $S_{\text{vN}} = -\operatorname{Tr}(\rho\ln\rho)$, where $\rho$ is the reduced density matrix of the QD. 
The entropy range $\ln 2 < S_{\text{vN}} < 2\ln 2$ reflects distinct physical regimes: the lower bound corresponds to a maximally entangled Kondo state, where the local spin forms a many-body singlet with itinerant electrons, while the upper bound represents an infinite-temperature thermal state with equal population across all Fock states.
Figure~\ref{fig1} shows the time evolution of various quantities at two reservoir temperatures: $\tilde{T} \equiv k_B T/\Gamma = 0.1$ (left panels) and $\tilde{T} = 1$ (right panels). 
The inset of Fig.~\ref{fig1}(b) demonstrates Kondo correlations at $\tilde{T} = 0.1$ through a prominent zero-frequency peak in the spectral function $A(\omega)$ of the equilibrium steady state. 
By contrast, $A(\omega=0)$ is strongly suppressed at $\tilde{T} = 1$ due to thermal fluctuations. 
We therefore refer to the $\tilde{T} = 0.1$ case as the correlated regime and $\tilde{T} = 1$ as the thermal regime.
Figures~\ref{fig1}(a) and (b) display the evolution of $S_{\text{vN}}$. 
Following the quench, $S_{\text{vN}}$ increases for both $\varrho_{\text{db}}$ and $\varrho_{\text{sm}}$ as correlations develop between the dot and electrode. 
The non-monotonic behavior of $S_{\text{vN}}(t)$ provides clear evidence of information backflow from the bath to the QD, characteristic of non-Markovian dynamics.

However, von Neumann entropy lacks sensitivity to directional information flow: 
Backflow can occur without a decrease in $S_E$ (e.g., when correlations suppress entropy reversal), and such backflow does not necessarily imply non-Markovianity. 
Non-monotonicity in the trace distance $D^{\text{tr}}$ provides a rigorous signature of information backflow and non-Markovianity \cite{PhysRevLett.103.210401,RevModPhys.88.021002}. 
Figures~\ref{fig1}(c) and (d) show the evolution of $D^{\text{tr}}$ for both correlated and thermal regimes. 
In the ideal correlated state, the dot-reservoir wave function forms a many-body singlet 
$
\ket{\psi}_{\text{tot}} = \frac{1}{\sqrt{2}} \left( \ket{\uparrow}\ket{\Downarrow} - \ket{\downarrow}\ket{\Uparrow} \right)
$, 
where $\ket{\Uparrow}$ and $\ket{\Downarrow}$ denote conduction electron spin states. 
The equilibrium steady state reduced density matrix is $\rho_{\text{ess}} = \frac{1}{2} \left( \ket{\uparrow}\bra{\uparrow} + \ket{\downarrow}\bra{\downarrow} \right)$. 
For this Kondo state, $D^{\text{tr}}(\varrho_{\text{db}},\rho_{\text{ess}}) = 1$ and $D^{\text{tr}}(\varrho_{\text{sm}},\rho_{\text{ess}}) = 0$. 
At infinite temperature, the thermal state distributes equally among all basis states, yielding $D^{\text{tr}}(\varrho_{\text{db}},\rho_{\text{ess}}) = 3/4$ and $D^{\text{tr}}(\varrho_{\text{sm}},\rho_{\text{ess}}) = 1/2$. 
Consequently, $D^{\text{tr}}$ for $\varrho_{\text{db}}$ exceeds that for $\varrho_{\text{sm}}$ in both regimes at $t=0$. 
Figures~\ref{fig1}(e) and (f) display the geodesic distance evolution, which exhibits dynamics similar to $D^{\text{tr}}$ (see Appendix~\ref{appb} for distance measure comparisons). 
In the correlated regime, $D^{\text{geo}}$ for $\varrho_{\text{db}}$ approaches equilibrium faster than for $\varrho_{\text{sm}}$ after a crossing time $\sim 10$ ps, demonstrating the presence of QME. 
This QME is absent in the thermal regime. 
Notably, $D^{\text{tr}}$ for $\varrho_{\text{sm}}$ in the correlated state shows multiple non-monotonic features, indicating strong non-Markovianity. 
The associated information backflow—manifested as temporary reversals of decoherence—delays relaxation toward equilibrium. 
This memory effect originates from the singly-occupied mixed state $\varrho_{\text{sm}}$: 
After an electron scatters off the impurity and flips its spin, the new spin state influences subsequent scattering events. 
This creates a feedback loop where current scattering outcomes depend on the history of spin-flip interactions. 
By contrast, the doubly-occupied state $\varrho_{\text{db}}$, with energy $U$ above the ground state, exhibits dominant occupation number relaxation and weaker spin-flip memory effects. 
The differential memory effects between initial states produce QME, a distinctly non-Markovian phenomenon. 
In the thermal regime [Fig.~\ref{fig1}(f)], both initial states exhibit similar decay dynamics without QME.

We now investigate the conditions for QME at fixed temperature. 
The QD exhibits distinct regimes parameterized by $\epsilon_d$: 
hole-type ($\epsilon_d \sim 0$) and particle-type ($\epsilon_d \sim -U$) mixed-valence regimes dominated by charge fluctuations, 
and the Kondo regime ($\epsilon_d \sim -U/2$) characterized by strong spin fluctuations. 
Figure~\ref{fig2} shows the evolution of $D^{\text{geo}}$ as $\epsilon_d$ varies from hole-type to particle-type mixed-valence regimes. 
In the hole-type mixed-valence regime ($\epsilon_d = 0$), both initial states exhibit exponential decay of $D^{\text{geo}}$ due to rapid charge fluctuations that prevent memory formation. 
At $\epsilon_d = -U/4$ (the crossover region between mixed-valence and Kondo regimes), memory effects emerge for $\varrho_{\text{sm}}$, as indicated by nonmonotonic $D^{\text{geo}}_{sm}$ behavior and slower decay. 
However, no crossing occurs between $D^{\text{geo}}_{\text{db}}$ and $D^{\text{geo}}_{\text{sm}}$. 
In the Kondo regime ($\epsilon_d = -U/2$), despite different short-time dynamics ($t < 20$ ps), 
$D^{\text{geo}}_{\text{db}}$ and $D^{\text{geo}}_{\text{sm}}$ converge for $t > 20$ ps. 
To quantify this criticality, we introduce the geodesic distance polarization 
$\chi \equiv (D^{\text{geo}}_{\text{db}} - D^{\text{geo}}_{\text{sm}})/(D^{\text{geo}}_{\text{db}} + D^{\text{geo}}_{\text{sm}})$. 
Figure~\ref{fig2}(f) shows $\chi$ as a function of $t$ and $\epsilon_d$. 
A narrow region with $\chi \sim 0$ appears for $\epsilon_d$ between $-U$ and $0$, indicated by the dashed line. 
The critical behavior at $\epsilon_d = -U/2$ arises from particle-hole symmetry, where the model is invariant under $d_s \rightarrow d_s^\dagger$. 
At $\epsilon_d = -3U/4$, the initial difference between geodesic distances decreases relative to $\epsilon_d = -U/4$ 
as the ESS is shifting to the doubly-occupied configuration. 
Combined with strong non-Markovianity in $D^{\text{geo}}_{\text{sm}}$, it results that $D^{\text{geo}}_{\text{db}}$ approaches to ESS slower than that of $D^{\text{geo}}_{\text{sm}}$, yielding QME (as illustrated by the asymptotic lines for long-time evolution in the inset of \Fig{fig2}(d)). 
In the hole-type mixed-valence regime ($\epsilon_d = -U$), $\varrho_{\text{db}}$ is closer to the ESS than $\varrho_{\text{sm}}$, 
and weak correlations preclude QME. 
These results demonstrate that QME occurs specifically in the crossover region between the Kondo and hole-type mixed-valence regimes.

\begin{figure}[htbp]
\centering
\includegraphics [width=1 \columnwidth]{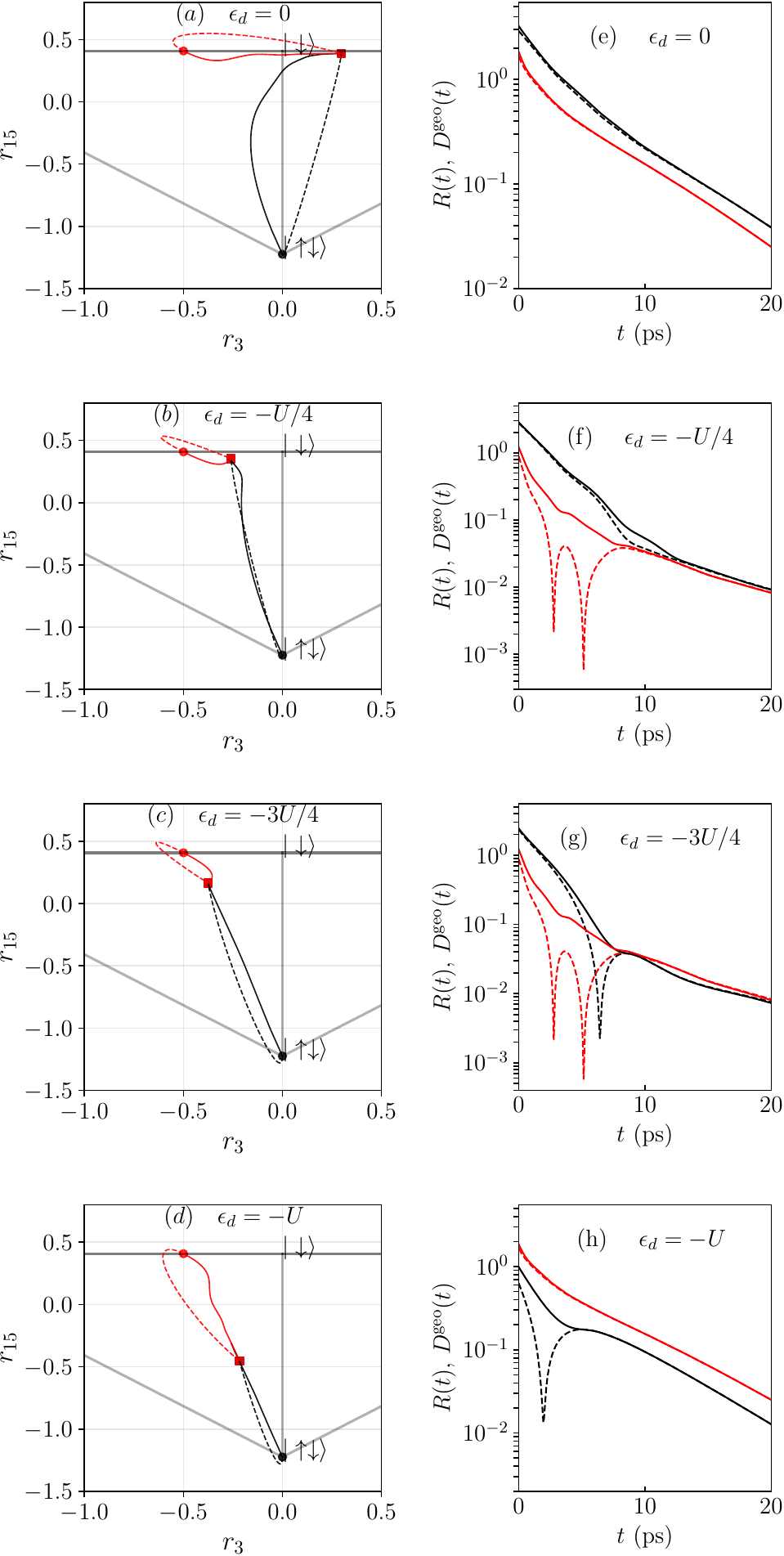}
\caption{(a)-(d) Evolution trajectories of $\rho_{\text{db}}(t)$ (black solid line) and $\rho_{\text{sm}}(t)$ (red solid line) for different $\epsilon_d$, where the initial states are marked by circles and the ESSs are marked by squares. The dashed lines are the  geodesics connecting the initial state and ESS.  
(e)-(h) The evolution of residue distances and geodesic distances corresponding to (a)-(d). Parameters are same as that in \Fig{fig2}.       }\label{fig3}
\end{figure}

Using the Bures metric for density matrices defined in Eq.~\eqref{b-metric}, we can construct the evolution trajectory under the completely positive trace-preserving (CPTP) map. 
The trajectory length reflects the intuition that a state geodesically closer to equilibrium may take longer to relax if it follows a more circuitous path \cite{qian2025intrinsic}. 
The length traced by the state as a function of time is given by
\begin{equation}
l(t)=\frac{1}{2} \int_{0}^{t}ds =\frac{1}{2} \int_{0}^{t}\sqrt{D(\rho,\partial_t \rho)}dt,
\end{equation}
in which
$
\sqrt{D(\rho,\partial_t \rho)} = \sum_{i,j|p_{i}+p_{j}\neq 0}\lvert \bra{i} \partial_t \rho\ket{j} \rvert^2 \frac{2}{p_i + p_j}.
$
The total trajectory length is $L = l(t \to \infty)$, and we define the residual distance as $R(t) = L - l(t)$. 
Unlike trajectory-independent distance measures, the intrinsic quantum Mpemba effect (IQME) is characterized using the trajectory length \cite{qian2025intrinsic}. 
Analogous to the QME criterion, IQME occurs when
\begin{equation}
\begin{aligned}
&R(\varrho_{A}) > R(\varrho_{B}), \\
\exists t_M > 0: \forall t > t_M: \quad &R(\rho_{A}(t)) < R(\rho_{B}(t)),
\end{aligned}
\end{equation}
indicating that state $A$, despite initially having a longer remaining trajectory, overtakes state $B$ after time $t_M$.

A general $4 \times 4$ reduced density matrix requires 15 real parameters for its description, corresponding to the generators of SU(4) \cite{KIMURA2003339,Bertlmann_2008,Bengtsson_Zyczkowski_2006}. 
In our model, mixing between different spin states is absent, as is mixing between different charge states due to charge conservation. 
Consequently, all off-diagonal elements vanish, leaving only the three parameters associated with diagonal generators. 
The density matrix expands as $\rho = \frac{1}{4}(I + r_3 \Lambda_3 + r_8\Lambda_8 + r_{15}\Lambda_{15})$, where $\Lambda_3$, $\Lambda_8$, and $\Lambda_{15}$ are diagonal generators of SU(4) (generalized Gell-Mann matrices). 
In the basis $\{\ket{0}, \ket{\uparrow}, \ket{\downarrow}, \ket{\uparrow\downarrow}\}$, these are defined as $\Lambda_3 = \operatorname{diag}(1,-1,0,0)$, $\Lambda_8 = \frac{1}{\sqrt{3}}\operatorname{diag}(1,1,-2,0)$, and $\Lambda_{15} = \frac{1}{\sqrt{6}}\operatorname{diag}(1,1,1,-3)$. 
The parameters $r_3$, $r_8$, and $r_{15}$ completely specify the state. 
The state space of diagonal density matrices can thus be visualized in three dimensions with coordinates $(r_3, r_8, r_{15})$, where each matrix corresponds to a point in $\mathbb{R}^3$ (see Appendix~\ref{appc}). 
Spin degeneracy imposes the constraint $r_8 = r_3/\sqrt{3}$, reducing the parameter space to two dimensions spanned by $(r_3, r_{15})$. 
Figures~\ref{fig3}(a)--(d) show trajectories for initial states $\varrho_{\text{db}}$ and $\varrho_{\text{sm}}$ at different $\epsilon_d$ (solid lines) in this generalized Bloch space, with circles (squares) marking initial states (ESSs). 
Dashed lines indicate geodesics connecting initial state and ESS. 
The evolution of $R(t)$ and $D^{\text{geo}}$ appears in Figs.~\ref{fig3}(e)--(h), where $D^{\text{geo}} \leq R(t)$ for all $t$ since the Bures geodesic is the shortest path under the metric in Eq.~\eqref{b-metric}. 
Notably, in the mixed-valence regimes ($\epsilon_d = 0$ and $\epsilon_d = -U$), $R(t)$ and $D^{\text{geo}}(t)$ coincide rapidly. 
By contrast, in the crossover regimes ($\epsilon_d = -U/4$ and $\epsilon_d = -3U/4$), convergence occurs much later, particularly for $\varrho_{\text{sm}}$. 
This behavior reflects distinct relaxation mechanisms: 
In mixed-valence regimes, $\epsilon_d$ lies near the Fermi energy, enabling strong charge fluctuations that drive rapid relaxation along nearly direct paths. 
In crossover regimes, Kondo correlations introduce slow many-body processes (e.g., spin flips), creating winding trajectories through intermediate states such as moment formation and quenching. 
Consequently, $R(t)$ deviates from $D^{\text{geo}}$ for extended periods, converging only near equilibrium. 
The intrinsic quantum Mpemba effect (IQME) emerges in the crossover regime, as evidenced by $R(t)$ curve crossings in Fig.~\ref{fig3}(g).

\textit{Conclusion-} In contrast to its  Markovian companion, identifying the non-Markovian QME presents significant challenges, as Lindblad eigenvalues alone cannot determine relaxation dynamics due to ubiquitous memory effects. 
We have demonstrated non-Markovian QME in quantum dots with Coulomb interactions, where Kondo many-body correlations generate strong memory effects that enable QME in the crossover region between Kondo and particle-type mixed-valence regimes. 
We anticipate that non-Markovian QME should occur in general strongly correlated open quantum systems under two conditions. 
First, the memory effect for initial state $B$ must significantly exceed that of state $A$, delaying the decay of $\rho_B$ despite its initial proximity to the equilibrium steady state. 
Second, the initial distance difference between $\rho_A$ and $\rho_B$ must be sufficiently small that the delayed relaxation of $\rho_B$ results in trajectory crossing.

\textit{Acknowledgments-} This work is supported  by the Natural Science Foundation of China through Grants No. 12404147.

\appendix
\section{Trace distance as measure for non-Markovian behavior }\label{appa}

Markovian dynamics are described by a divisible completely positive trace-preserving (CPTP) map $\Phi_t = \Phi_{t,s} \circ \Phi_{s}$ for $t>s$, which satisfies
\begin{equation}
D^{\text{tr}}(\Phi_t(\rho),\Phi_{t}(\sigma)) \leq D^{\text{tr}}(\Phi_{s}(\rho), \Phi_{s}(\sigma)).
\end{equation}
This monotonic decrease of the trace distance reflects irreversible information flow to the environment. 
For non-Markovian dynamics, the overall evolution map $\Phi_{t}$ (from $0$ to $t$) remains CPTP, so
\begin{equation}\label{cptp}
D^{\text{tr}}(\Phi_t(\rho),\Phi_{t}(\sigma)) \leq D^{\text{tr}}(\rho_0, \sigma_0),
\end{equation}
preserving contractivity for fixed $t$. 
However, the intermediate map $\Phi_{t,s}$ (from $s$ to $t$) may not be CPTP, allowing
\begin{equation}
D^{\text{tr}}(\Phi_t(\rho),\Phi_{t}(\sigma)) \geq D^{\text{tr}}(\Phi_{s}(\rho), \Phi_{s}(\sigma))
\end{equation}
for some $t \geq s$. 
This temporary increase signifies information backflow from the environment, distinguishing non-Markovian dynamics while maintaining the contractivity principle. 
The assumption of initially uncorrelated system and environment is essential for ensuring that $\Phi_t$ is CPTP. 
Otherwise, Eq.~\eqref{cptp} can be violated when using correlated or entangled system-reservoir states, as the system's evolution depends on hidden correlations that break consistency under arbitrary extensions \cite{Laine_2010,PhysRevA.82.012341}. 
Consequently, initial system-reservoir correlations can cause transient increases in the distance from equilibrium. 
Recent work has demonstrated QME induced by initial system-reservoir entanglement \cite{10.1063/5.0266143}. 
To isolate non-Markovianity as the sole origin of temporary trace distance increases, we therefore adopt uncorrelated initial states.

\section{Different choices to quantify the distance from equilibrium}\label{appb}

\begin{figure}[htbp]
\centering
\includegraphics [width=1 \columnwidth]{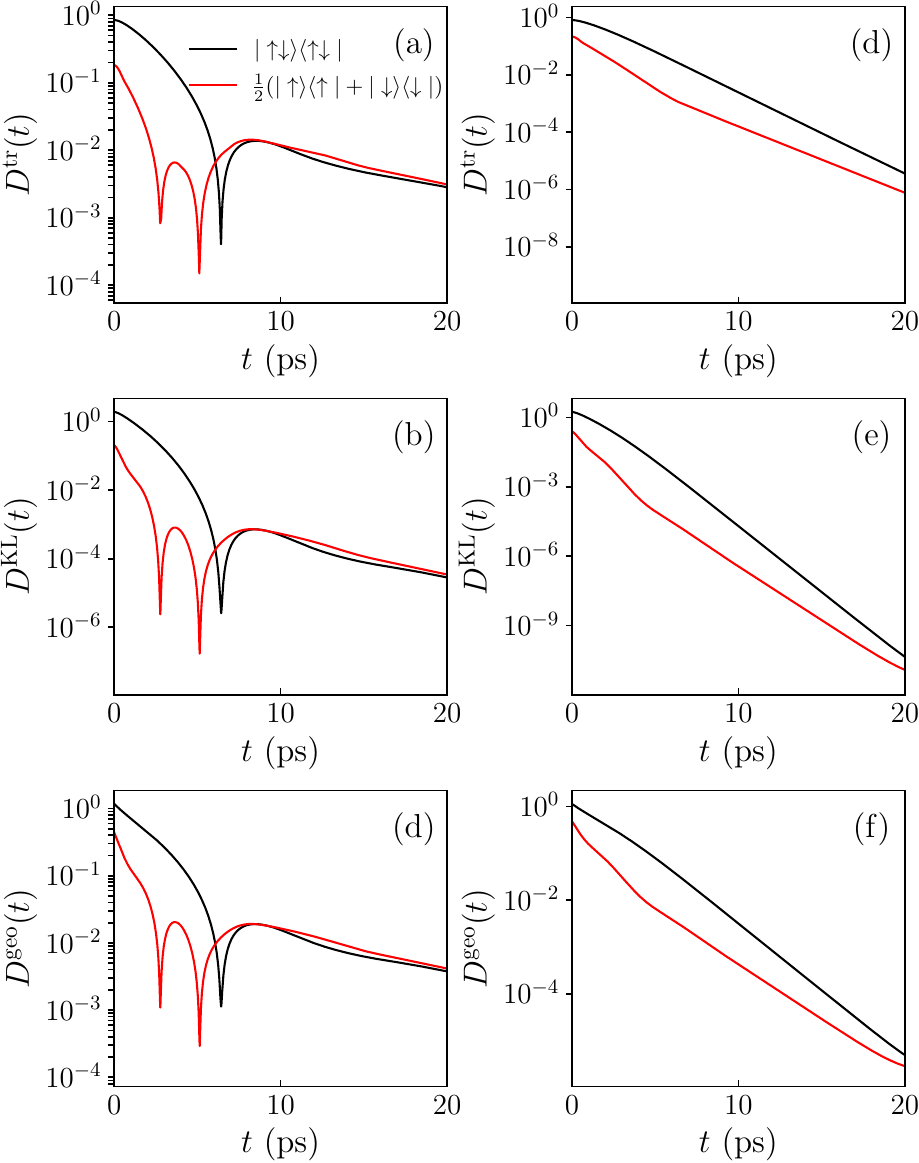}
\caption{(a) and (d) Trace distance,  (b) and (e) quantum relative entropy, (c) and (f) geodesic distance evolution for different initial states. Temperature for the left (right) panel is $\tilde{T}=0.1$ ($\tilde{T}=1$). The single electron energy level is $\epsilon_d = -3U/4$, other parameters are same as that in \Fig{fig1}.  
  }\label{sm-1}
\end{figure}

Multiple measures can quantify a system's deviation from equilibrium. 
Beyond trace distance and geodesic distance, the quantum relative entropy between $\rho(t)$ and $\rho_{\text{ess}}$ also measures state distinguishability. 
This quantity, related to nonequilibrium free energy, provides another candidate for defining QME in open quantum systems \cite{PhysRevLett.133.140404}. 
The quantum relative entropy, also known as the Kullback-Leibler (KL) divergence, is defined as
\begin{equation}
D^{\text{KL}}(\rho,\sigma) = \operatorname{Tr}[\rho(\ln\rho - \ln\sigma)].
\end{equation}
It captures optimal state distinguishability in a single measurement and upper-bounds the trace distance via Pinsker's inequality:
\begin{equation}
2[D^{\text{tr}}(\rho,\sigma)]^2 \leq D^{\text{KL}}(\rho,\sigma).
\end{equation} 
Relationships between these distance measures include the trace-geodesic inequality
\begin{equation}
1 - \cos D^{\text{geo}} \leq D^{\text{tr}} \leq \sqrt{1 - \cos^2 D^{\text{geo}}},
\end{equation}
and the relative entropy-geodesic inequality
\begin{equation}
2\ln(\cos D^{\text{geo}}) \leq D^{\text{KL}}.
\end{equation}
Figure~\ref{sm-1} compares these distance measures for the crossover region at $\epsilon_d = -3U/4$. 
Despite differences in magnitude and line shape, all measures consistently indicate QME presence for correlated regime at low temperature and absence for thermal regime at high temperature.

\section{geodesics in Bloch parameter space}\label{appc}

For diagonal density matrices $\rho = \operatorname{diag}(\bm{\varrho})$ and $\sigma = \operatorname{diag}(\bm{\varsigma})$, the Bures distance simplifies to a function of the Bhattacharyya coefficient $B(\varrho,\varsigma)$:
\begin{equation}
D^B = \sqrt{2\left(1 - \sum_i \sqrt{\varrho_i \varsigma_i}\right)} = \sqrt{2(1 - B(\varrho,\varsigma))},
\end{equation}
which corresponds to the Euclidean distance between points on a unit sphere in square-root probability coordinates. 
The Bures angle $D^{\text{geo}} = 2 \arccos\left[1 - (D^B)^2/2\right]$ gives the geodesic distance along this sphere. 
For diagonal density matrices, the line element of the Bures metric in Eq.~\eqref{b-metric} reduces to
\begin{equation}
ds^2 = \frac{1}{4} \sum_{i=1}^{4} \frac{dp_i^2}{p_i}.
\end{equation}
Considering a diagonal density matrix $\rho = \operatorname{diag}(p_1, p_2, p_3, p_4)$ in the basis $\{\ket{0}, \ket{\uparrow}, \ket{\downarrow}, \ket{\uparrow\downarrow}\}$, the Bloch parameters are defined as
\begin{equation}
\begin{aligned}
r_3 &= p_1 - p_2, \\
r_8 &= \frac{p_1 + p_2 - 2p_3}{\sqrt{3}}, \\
r_{15} &= \frac{p_1 + p_2 + p_3 - 3p_4}{\sqrt{6}},
\end{aligned}
\end{equation}
with the inverse relations
\begin{equation}\label{probs}
\begin{aligned}
p_1 &= \frac{1}{4}\left(1 + r_3 + \frac{r_8}{\sqrt{3}} + \frac{r_{15}}{\sqrt{6}}\right), \\
p_2 &= \frac{1}{4}\left(1 - r_3 + \frac{r_8}{\sqrt{3}} + \frac{r_{15}}{\sqrt{6}}\right), \\
p_3 &= \frac{1}{4}\left(1 - \frac{2r_8}{\sqrt{3}} + \frac{r_{15}}{\sqrt{6}}\right), \\
p_4 &= \frac{1}{4}\left(1 - \frac{3r_{15}}{\sqrt{6}}\right).
\end{aligned}
\end{equation} 
The constraints $p_i > 0$ define a physical parameter space forming a tetrahedron in $(r_3, r_8, r_{15})$ coordinates. 
The vertices correspond to pure states:
\begin{equation}
\begin{aligned}
\ket{0}&: \, (r_3, r_8, r_{15}) = \left(2, \frac{2}{\sqrt{3}}, \frac{\sqrt{6}}{3}\right), \\
\ket{\uparrow}&: \, (r_3, r_8, r_{15}) = \left(-2, \frac{2}{\sqrt{3}}, \frac{\sqrt{6}}{3}\right), \\
\ket{\downarrow}&: \, (r_3, r_8, r_{15}) = \left(0, -\frac{4}{\sqrt{3}}, \frac{\sqrt{6}}{3}\right), \\
\ket{\uparrow\downarrow}&: \, (r_3, r_8, r_{15}) = \left(0, 0, -\frac{2\sqrt{6}}{3}\right),
\end{aligned}
\end{equation}
while mixed states occupy the interior. 
The tetrahedron center corresponds to the maximally mixed state with $p_i = 1/4$. 
For arbitrary Hilbert space dimensions, the Bures geodesic between $\rho$ and $\sigma$ is given by \cite{Ericsson_2005}
\begin{equation}
\begin{aligned}
\lambda(\tau) =& \frac{1}{\sin^2 \theta} \bigg[ \sin^2(\theta(1-\tau)) \rho + \sin^2(\theta \tau) \sigma \\
&+ \sin(\theta(1-\tau)) \sin(\theta \tau) \left( \rho^{-1/2} \lvert \sqrt{\sigma} \sqrt{\rho} \rvert + \text{h.c.} \right) \bigg],
\end{aligned}
\end{equation}
where $\theta = \arccos \mathcal{F}(\rho,\sigma) = D^{\text{geo}}/2$ and $0 \leq \tau \leq 1$. 
In our quantum dot system, spin degeneracy imposes $p_2 = p_3$. 
From Eq.~\eqref{probs}, this constraint yields $r_3 = \sqrt{3} r_8$, reducing the parameter space from three dimensions $(r_3, r_8, r_{15})$ to two dimensions $(r_3, r_{15})$.

\begin{figure}[htbp]
\centering
\includegraphics [width=1 \columnwidth]{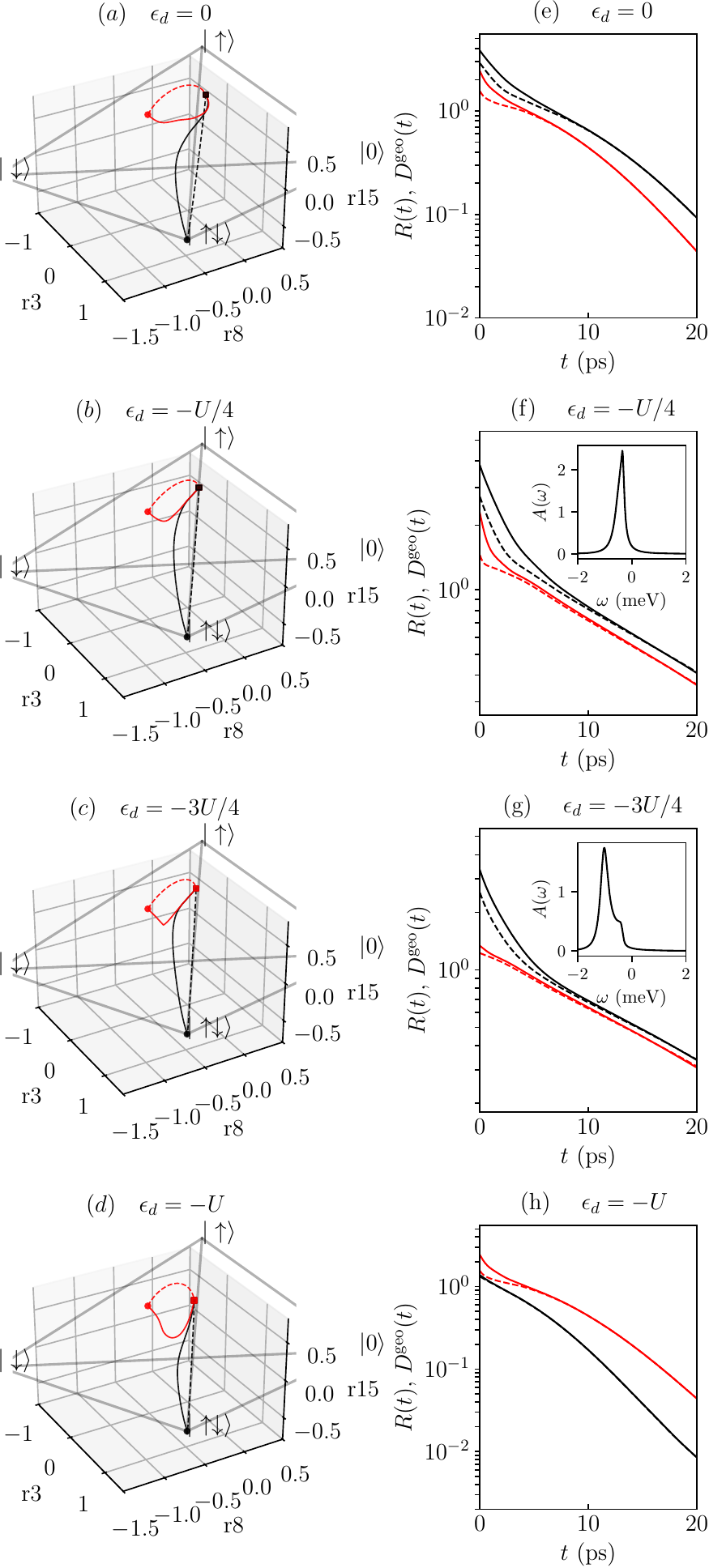}
\caption{(a)-(d) Evolution trajectories of $\rho_{\text{db}}(t)$ (black solid line) and $\rho_{\text{sm}}(t)$ (red solid line) for different $\epsilon_d$ in generalized Bloch space, where the initial states are marked by circles and the ESSs are marked by squares. The dashed lines are the  geodesics connecting the initial and equilibrium states. The corners of the tetrahedron are the four pure Fock sates. (e)-(h) The evolution of residue distances and geodesic distances corresponding to (a)-(d). Spin splitting is set as $\Delta_s/\Gamma = 1$,  other parameters are same as that in \Fig{fig3}.      
  }\label{sm-2}
\end{figure}

Spin degeneracy can be broken by applying a magnetic field to the quantum dot, described by the Hamiltonian
\begin{equation}
H_{\text{dot}} = \sum_{s} \epsilon_{d,s} d_{s}^{\dagger} d_{s} + U n_{\uparrow} n_{\downarrow},
\end{equation}
where $\epsilon_{d,s} = \epsilon_d - \Delta_{s}$, with $\Delta_{s} \equiv \frac{1}{2} s g\mu_B B$ and $s = \pm 1$ denoting spin orientation. 
Figures~\ref{sm-2}(a)--(d) show trajectory evolution in the generalized 3D Bloch parameter space for different $\epsilon_d$ values with $\Delta_s / \Gamma = 1$. 
The corresponding evolution of $R(t)$ and $D^{\text{geo}}$ appears in Figs.~\ref{sm-2}(e)--(h). 
Kondo singlet formation requires energy degeneracy between the dot electron and reservoir electrons. 
The magnetic field-induced Zeeman splitting breaks this degeneracy, making singlet formation energetically unfavorable and suppressing Kondo correlations. 
As shown in the insets of Figs.~\ref{sm-2}(f) and (g), the Kondo peak in $A(\omega=0)$ is suppressed for both $\epsilon_d = -U/4$ and $\epsilon_d = -3U/4$. 
In contrast to the zero-field case discussed in the main text, non-Markovian features are absent from the evolution dynamics regardless of $\epsilon_d$ (right panels of Fig.~\ref{sm-2}). 
Consequently, both QME and IQME are absent, as evidenced by the lack of crossings in $R(t)$ and $D^{\text{geo}}(t)$ curves.

\bibliography{Refs}

\end{document}